# Milestones for Teaching the Spreadsheet Program

Étienne Vandeput

CRIFA – STE – FAPSE – ULg, boulevard du Rectorat 5, B – 5000 Liège (Belgium)

evandeput@ulg.ac.be

**ABSTRACT**

*There are different manners of teaching a spreadsheet program. In any case, it is intended that the teacher settles the objectives of the course and adapts them to the particular audience he/she has to deal with. This paper aims at providing any teacher whatever his/her specific objectives and his/her audience with elements to help him/her building a course. It focuses mainly on two important issues: 1 - select in all that may be said about such complex tools, what is prior to know and to teach, i.e. what leads to autonomy in using but also to autonomy in learning (because everything cannot be taught) and 2 - show how concepts are closely related to good formatting considerations. A method based on the "invariants of information processing" is outlined, partially illustrated and an implementation is described throughout a course designed for students preparing a master in Education Sciences.*

## 1  INTRODUCTION

Talks around « must one teach something, and if so, what about office automation tools? » have been dealt with for years. In Belgian secondary schools, by the end of the eighties, many teachers decided to renounce teaching programming because it seemed too hard. Most of them found that teaching about office automation tools with ICT was more interesting for students and more in relationship with what they would be asked to do later. They were right and wrong at the same time. Of course, teaching the art of programming is not so easy. It is probably the main reason why they gave it up. But they got the feeling that teaching office automation tools was simple and limited to menus exploration as well as to exercises with an online help or a user's guide consultation. We never thought so.

On the other hand, a spreadsheet program is more complex than a word processing program or a presentation program because processing still remains more formal, even generic and allows applications in a very wide range of domains. So, we do not have any reason to say that "There is nothing to teach about using a spreadsheet program.".

In its 2$^{nd}$ section, this paper proposes to define what general objectives of a course on the spreadsheet program should be and to point out different elements that cannot be ignored to reach such objectives. The definition of the milestones of a course on the spreadsheet program is made easier using a method based on the "invariants of information processing". The method is explained and partially implemented. Further, in section 3, we illustrate an implementation of these steps in a course sequence dedicated to students preparing a master in education sciences. A majority of them will need to use a spreadsheet program in the framework of his/her master's dissertation. Section 4 suggests a conclusion giving a feedback on the method developed in section 2 and guidelines inspired by the examples shown in section 3.

## 2 THEORETICAL REFLECTIONS

### 2.1 Short Considerations About Learning

To begin with, we think that some learning processes should be compared. Nowadays, teachers may use a range of strategies and tools to build their course scenario. E-learning, blended learning and so on provide them with a variety of environments and contexts they can use.

Of course, there are many ways leading to learning. And the depth of learning may also vary. Using a tutorial does not mean training, and training is not really the same as teaching. However, ICT didactical and pedagogical considerations are linked to such ambiguities. Many people (and unfortunately also teachers and trainers) sometimes consider teaching, training and using a tutorial as identical things. In this section, we will quickly try to explain what differences can be identified between these three ways of "making learn". In the following sections, we will insist on what steps can be considered as fundamental in case teaching is the chosen way to make learn.

**Tutorials**

ICT learning often consists in ICT e-learning. In fact, if ICT exist, why don't we use them to learn, and particularly to learn about them? Teachers and trainers have the opportunity of creating learning supports like CDs, web sites, web applications... including various multimedia items like pictures, videos, animations, screen casts... That is what we could call "the big temptation". The process becomes attractive and many people think that motivation grows when the learner is sitting in front of a computer screen. They are probably right. But, what about learning?

Using such a system makes it easy to « show ». So, the teacher (designer) can center the learning on the impregnation/modeling paradigm but also provide the learner with some pieces of advice developing the practice/guidance paradigm [Denis & Leclercq, 1994]. Though it is not a general rule, a reflection process is often missing because a tutorial generally emphasizes demonstrative aspects. The trainer entrusts the system with his/her own role, losing part of the interaction richness.

Notice that: tutorials, screen casts and other similar tools are often more useful when they are included in a more global pedagogical scenario initiated by a teacher.

**Training**

Even when the learning is not from a distance, the use of software in order to carry out a task is often considered like a practical process devoid of any intelligent approach. This point of view is unfortunately shared by a majority of learners and teachers. For instance, the use of a word processing program is considered by lots of people like a sequence of elementary commands. Such an approach cannot lead to a structured and efficient work. Nevertheless, many trainers build their training on this methodology. What the learner can see remains very important. So the trainer will insist on the graphical elements of the environment (menus, buttons, checkboxes…) and on graphical aspects of the process results. Examples about the spreadsheet are numerous. For instance, trainees learn to enlarge a column or to emphasize characters before writing a formula or thinking about the modeling of a spreadsheet. They are very surprised when a value does not appear in a cell like they thought it would, but they do not care about cell format and information types. They feel very satisfied when they get a pretty graphic even if it shows wrong values and so on.

Vocational training often has precise and operational objectives and aims at developing professional skills. Sometimes, the trainees have to learn special routines related to their current or future job or activity. So, reflecting about the deep nature of things is not always a prior concern. That is why many trainers often hesitate to build their training on long-term effects and insist more on practical results rather than on competences gathering knowledge and know-how.

**Teaching**

If training often refers to a learning activity parallel to or following studies (sometimes from afar), with very precise objectives and sometimes expressed in terms of technical know-how, teaching rather refers to studies leading to a diploma. The objectives for the learners are related to more global competences. That means a much more long-term view. Skills to be developed are identified, but the processes aren't especially determined. Of course, this paper is dedicated to teaching in an academic context. So, let us develop in the next sections our point of view on teaching.

**2.2 Objectives of a course on the spreadsheet program**

Regardless of any audience, one can agree on a very general objective: *make and keep the learners autonomous in the use but also in the learning of a spreadsheet program*.

What does that mean?

The first part is easier to understand than the second one. In fact, the learner should offer proof that he/she is able to summon up his/her knowledge of the spreadsheet program to solve all kinds of problems related to the management of spreadsheets: building, modifying, correcting... Such an objective can easily be detailed into more operational ones. It is not so complicated to identify which knowledge (a small part depending on the audience) should be mastered. Trails are provided in the next section.

The second objective is more difficult to reach. The spreadsheet program, like many other ones, is very complex. It is well known that many people use only a few features among those that are offered. Some of these features are sometimes so specific that users need them in very particular cases. So, for a user (learner), it's less important to know a lot about a software package than to be able to discover a feature when needed, because he/she has got good representations of how the global task can been formalized and how a system works. In other words, providing learners with such good representations (if needed) is quite essential. We find particularly important to insist on the formal size of numerical processing. It helps the learner to understand and sometimes to imagine how the system works or could work. We have developed this point of view in [Vandeput, 2006]. Some examples are given in the experience developed at the end of the present paper. These clearly illustrate that fundamental constraint due to any computer-based systems.

Let us come back to the first objective. There are two sub-objectives to reach and they are complementary. The first one is to master the basic concepts. Notice that *basic* does not mean *simple*! The second one is to be able to format a spreadsheet efficiently. Of course, this efficiency concerns the flexibility of the template (*e.g.* What if that data is modified?

Does the spreadsheet fit with changes?), but also the quality of information presentation (special data, results, graphics...).

The first is prior to the second. A learner will be unable to format a spreadsheet if he/she does not master the main concepts. As you will see in the final illustration, it does not mean formatting should be moved to the end of the course. Formatting is a know-how that can be developed at any stage of the learning.

To identify the basic concepts, we use a method which is applicable to any software package formalizing a task. It is based on what we call "invariants of information processing". This method is applied to the spreadsheet program in the following section.

**2.3 Method to elicit basic knowledge**

**Invariants of information processing**

The method and its advantages are developed in [Poisseroux & al., 2008]. We shortly provide some explanations about it.

Obviously, the term "invariant" is used to call "what does not vary, what stays constant, fixed or stable". In the domain of information and communication technologies (ICT), identifying invariants is useful. Users can focus on their goals (how to reach them efficiently) and not on what they can see (graphical environment). This way of thinking makes them become independent of the software interface and further, more autonomous using, but also learning about software.

Invariants can be either global, or related to specific software (e.g. modeling a task such as mail, document edition, information retrieval, management...). They may have two different natures. They are either concepts or (tasks) organizing principles. An organizing principle describes an automatic or semi-automatic process carried out by the system (any association "computer-software" at a particular moment of a work session) and dedicated to a task (*e.g.* word processing) or a set of tasks (*e.g.* office automation). An organizing principle is a specific software feature intended to shorten the user's task. Here are some examples.

Automatic calculation is a basic and very important organizing principle of the spreadsheet program. Power typing in word processing is another good example. Many programs implement automatic data capture (browser, spreadsheet, email…). The user does not need to encode long character strings (URL, data in a cell, email address…).

Organizing principles are invariants by nature because they are closely related to the task that does not fundamentally -or so slowly- evolve. Here are some examples. It sounds like an evidence that a spreadsheet implements automatic calculation or that word processing implements word wrap. Principles are stated using concepts. Because of this relation, concepts must be qualified as invariants. A paragraph (document edition), a slide (presentation), a cell, a formula (spreadsheet)... are concepts adopted by most of the software developers.

From a pedagogical point of view, invariants are pieces of knowledge. Knowing them should have a good influence on the user know-how, helping him/her to manage efficiently the system.

Like for other pieces of software, the spreadsheet developer has to formalize a task. Quickly and simply said, he/she has to imagine what kinds of goals the user will formulate when carrying out this task.

4Copyright © 2009 EuSpRIG & The Author(s)

**Spreadsheet program invariants**

As a strategy, we could decide to identify a large number of invariants and to classify them into different categories. This has been done in [Vandeput & Colinet, 2005]. But there is another way to use invariants, that is closer to a didactical approach. That consists in:

1. starting from the more important organizing principles (the more related to the task, the more useful from an information processing point of view, the software core activity);
2. deriving the main concepts linked to these principles;
3. identifying other principles and concepts inspired by these concepts;
4. iterating steps 2 and 3.

This way of thinking helps defining a framework to build a course. Indeed, one important question when giving a course whatever it is, and this issue is particularly hard in the ICT domains (where things to be known have not been filtered and classified), is to determine which concept should be introduced first, and what about the next one, and the following one... Here is how we could proceed with the spreadsheet program.

As a matter of fact, the main principle organizing work with the spreadsheet is the *recalculation principle*, responsible for a spread of changes. It could be formulated as follows:

"The system (computer + spreadsheet program + OS of course) is able to **automatically recalculate** any value (piece of information) contained in a *cell* and depending, through a *formula*, on other values (pieces of information) contained in other cells of the same sheet, the same *spreadsheet* or possibly other spreadsheets".

This principle should be defined more precisely but it contains all we need to explore the first invariant concepts related to it. Notice that each principle can be formulated in a sentence starting by "The system is able to **automatically…**"

Some concepts are entities: *cell*, *sheet*, *spreadsheet*... Most of software packages are based on entities handlings: *word*, *paragraph*, *page*... (word processing program), *slide*, *title*,... (presentation program), *cell*, *column*, *row*, *sheet*,... (spreadsheet program). Early identifying those entities is necessary because they are important pieces of the puzzle. The learner will have to know what kind of features may be applied to them and what are their attributes. Of course, entities are invariant concepts. Every spreadsheet program uses them in the same way. Some short differences may be observed as far as the features and/or the software instructions are concerned but not on the entities and how they are named.

Another important concept mentioned in the statement of the principle is the *formula*. When thinking about what is a formula, new concepts appear. Reflecting upon what makes up a formula, the response is very simple. A formula is an expression including some of the following elements: *operators*, *constant* values, (possibly) *variable* values and/or *functions*.

*Values* and *variable* make one think about *types* but *variable* also introduces the *reference principle*.

Notice that only four kinds of elements are used to build a formula. So, it is not complicated to understand the components of the formula and to to identify them easily. Identically, there are lots of functions but the learner does not need to know all of them. He/she must know the general syntax of a function. So, new concepts appear such as, *identifier* (of the function), *arguments*, *separators*...

The *type* concept evokes the very transversal *coding principle* that can be defined like this: "The system is able to code (digitize) information depending on what kind of operations will be applied on it and to decode digitized information so that it can be displayed.". Notice that few types exist and that they can be grasped at the same time.

The *type* concept is inseparable from the *formatting principle*. One can say: "The system is able to display information in a default format or any other format defined and/or indicated by the user."

At this stage we reach a kind of "Pons Asinorum". Indeed, if the user (learner) encodes data before any formatting step, the spreadsheet program chooses by itself the type, the display format and the layout format, which can sometimes be very amazing. So, all these concepts must be clearly understood.

Once the *reference principle is introduced*, another principle arises: the *(re)copy principle*. This principle is connected to the following concepts: *absolute reference*, *relative reference* and *reference by name*.

That quick brainstorming shows how principles may evoke concepts and concepts may evoke other principles. A course sequence can be based on such a brainstorming. We recommend to provide illustrations and to make exercises introducing these concepts and principles one by one.

**Good Practice Guidelines**

The building of a « well-formatted» spreadsheet is related to a good knowledge of the invariants. For instance, the *(re)copy principle* has an influence on the spreadsheet design.

Thinking about the *format principle*, the good designer will determine him/herself the type and display format for all the significant parts of the sheets.

Another principle like the *conditional formatting principle* is also a tool to exploit in order to put some data in evidence and increase the quality of data presentation.

In the following sections, we develop and illustrate the didactical approach mixing the discovery of spreadsheet program invariants and the drawing up of guidelines.

## 3 CASE OF STUDENTS IN EDUCATION SCIENCES

### 3.1 Context Description

At the University of Liege (ULg - Belgium), teachers, social and youth workers are given the opportunity of applying for a master in Educational Sciences. But, in order to get the access to the master without coming from a bachelor degree in the domain, they first have to follow a « one-year preliminary program leading to the Master in Educational Sciences ». This foundation course is composed of different courses aiming at providing them with some necessary basis. One of these courses is called « Introduction to the computerized processing of quantitative and qualitative data ». It should, in the future, help students analyzing their data with, for instance, making reports or their master's dissertation. As it is a flexible and performing tool, an important part of this course is dedicated to the spreadsheet program.

Let us characterize the audience more precisely. Students are preparing a Master in Educational Sciences. Next year, they will have to choose a master's dissertation topic. This dissertation is, most of the time, an exploratory survey needing data analysis, graphical representations and related comments. That is why they often need to use a spreadsheet program. Most of them have a very short experience in the use of this kind of program. On top of that, some of them do not have any idea of what a spreadsheet program can be. The course aims at providing them, quickly, with good bases to lead such a work efficiently.

**3.2 Didactical Approach**

« What should be taught, and how? » is a relevant and certainly not a trivial question.

The objective of the course is very clear. Students should demonstrate their efficient use of a spreadsheet program in order to analyze, graphically present, and implement formulas to data. The duration of the course is 20 hours including theory and practical work. Time is short. So, we have to think about a didactical trail adapted to the situation.

This section describes how activities like teaching concepts (those that have been quickly evoked in the previous section), providing guidelines, carrying out formative evaluation are mixed in order to achieve the objective mentioned.

In order to make the learning process shorter and more efficient, we focus on two concerns:
- the course content (what students must absolutely know about a spreadsheet program);
- the art of building a spreadsheet.

Both concerns may (must) be dealt with simultaneously. The second point, of course, focuses on the main recommendations in relationship with the tasks that the students are likely to realize in the future. The teacher will find a very complete inspiration source in [Read & Batson, 1999].

Here are, accompanied with some comments, ten steps of the didactical trail. The pedagogical scenario is very simple. It includes, for each trail, an alternation of theoretical discussions and practical sessions.

**First Step: Computer-Based Systems are Processing Information Formally.**

Because of the low knowledge level of the students (a quick survey provided this information), the step is unavoidable. A session is devoted to making them understand well how a numerical system works. At the end, students must have understood that « in order to be processed by a computer, information must be digitized ». They should become familiar with coding and understand that coding has an influence on processings that, as a consequence, always concern information format and never its meaning.

Various situations are described. Here are some examples.

Same meaning but different formats:

Copyright © 2009 EuSpRIG & The Author(s)                                            7

*Six, SIX, six, 6, VI...*
*Computer, COMPUTER, ComPutER, computter...*

Same format but different meanings:
...a *train* to Paris..., ...my *train* of thought..., ...to *train* or to teach...

Sometimes, frequent problems are used to underline the importance of coding like in the following example (extract from a French email message including stressed characters)

*Bonjour Avant tout une bonne ann=E9e 2007 =E0 toutes et =E0 tous. Ceci =E9tant dit, passons aux choses p=E9nibles. Les accords de d=E9cembre pass=E9s par Marie Arena limitent les heures "non enseign=E9es".*

**Second Step: Formatting of a Spreadsheet from an Existing Document Starts with an "Atomization" of Information.**

[Invoice image: Proximus Belgacom Mobile invoice, N° DE FACTURE: 010903689211, Compte Client: 04597057, Date: 22 octobre 2001. Client JEAN, facturation JEROME. RESUME Pour la période: 10 sept. 2001 - 09 oct. 2001. Abonnement 21% 10,7556; Communications 21% 313,4807. Tarif T.V.A. 21% Total en EUR; Montant de base 324,24 324,24; Montant T.V.A. 68,09 68,09. Montant total de cette facture: 392,33. A payer avant le 06 novembre 2001 392,33. (Total équivalent en BEF: 15.827)]

Just an example to show that converting an existing document into a table is not so trivial.

The framework rows-columns does not appear. But another issue is to know how information must be broken down like for instance: « A payer avant le 06 novembre 2001 ». One cell, two cells ore more?

Here is a first opportunity to speak about good formatting.

**Third Step: the User (Learner) Should Have the Control, not the System.**

Exercises are proposed to show the learners that the system must be controlled. They are invited to put some particular character strings in the cells of a sheet and have to understand why the system displays something different from what they were thinking about.

Here are such strings. You can introduce them, one by one, in the cells of a sheet. What is really displayed is often amazing. What happens? The explanation is not always very simple.

*123     12/3     123,4567        12-3     12+3     12/12/12          63%     12:3     12/15*
*22/22   29/2/4   29/2     13:63     25:30     12/12 6:20          12/22 6:00*

**Fourth Step: the User Controls but the System Helps (a Lot).**

Before speaking about formulas, generation of series is a good example of how the system helps the user (learner). What happens when expanding down the cells B2, C2..., F2 and what about expanding the blocks G2..G3, H2..H3 and I2..I3 ?

|   | A | B | C | D | E | F | G | H | I | J |
|---|---|---|---|---|---|---|---|---|---|---|
| 1 |   |   |   |   |   |   |   |   |   |   |
| 2 |   | Trim 1 | Week 1 | March-2000 | John Lennon |   | 8:00 | 8:00 | 1 | 1,5 |
| 3 |   |   |   |   |   |   |   | 8:10 | 2 | 1,6 |
| 4 |   |   |   |   |   |   |   |   |   |   |

Good formatting of a spreadsheet takes these considerations into account.

**Fifth Step: Formulas Are the Core Business of the Spreadsheet Program.**

The first examples and exercises focus on the identification of calculated values. Formulas do not include functions. The table is adapted to the discipline because it is related to pre- and post-tests. At this stage, the (re)copy feature is not necessary but may be used.

|   | A | B | C | D | E |
|---|---|---|---|---|---|
| 1 |   | PRE | POST | BENEFIT | R. B. |
| 2 | *Vocabulary* | 33% | 70% | 37% | 55% |
| 3 | *Grammar* | 63% | 86% | 23% | 62% |
| 4 |   |   |   |   |   |

**Sixth Step: (re)Copy Is a Basic Tool for the Good Formatting of a Spreadsheet.**

For instance, how many different formulas are needed to build this table?

|   | M | N | O | P | Q |
|---|---|---|---|---|---|
|   | * | 6 | 9 | 8 | 2 |
|   | 5 | 30 | 45 | 40 | 10 |
|   | 7 | 42 | 63 | 56 | 14 |
|   | 9 | 54 | 81 | 72 | 18 |
|   | 3 | 18 | 27 | 24 | 6 |

Just one. And which one?

**Seventh Step: Other Basic Tools Are Cells Blocks Formatting and Conditional Formatting.**

The following example shows that particular exercises may motivate learners and help them learning more and by themselves. In that case, an exercise makes sense, even if it is not directly related to the learner's discipline.

|   | A | B | C | D | E |
|---|---|---|---|---|---|
| 1 |   |   |   |   |   |
| 2 |   | apple | pear | banana | cherry |
| 3 |   | grapefruit | apricot | pineapple | hazelnut |
| 4 |   | melon | orange | lemon | watermelon |
| 5 |   |   |   |   |   |
| 6 |   |   |   |   |   |

Any fruit name, whatever it is, written in this table, must be displayed in red if the name is more than six characters long.

Students like fruits but also this kind of exercise.

**Eighth Step: the Spreadsheet Program Is Helpful When Data Are Numerous.**

This step is probably a bit more specific to the audience. Surveys they have to do often concern lots of people, lots of questions or items... To teach them how to modify or simply consult large tables (hiding rows or columns, using elaborated filters or not...) is certainly a mean to increase their efficiency.

**Ninth Step: Data Sometimes Already Exist Elsewhere.**

This step is also quite specific to the audience. It helps increase efficiency. As data often exist elsewhere (in another sheet created by another spreadsheet program, in a table of a text document or if they are structured in a text document...) they may easily be imported.
Notice that, more generally, exercises about these features prevent students from losing time encoding and allow them to focus on the important thing. So, for instance, they can work on large data sets (see previous steps).

**Tenth Step: the System Easily Generates Charts but, Designing a Good Chart Is not so Easy.**

Concepts about charts are limited. Learners quickly discover how to ask the system to display a chart. In this case, sobriety is often missing. But choosing a relevant type, avoiding redundancy, choosing scales in order to provide a good idea of what data should show... are abilities that should be promoted.

**4   CONCLUSION**

It is not so obvious to point out what is important when teaching about ICT. The reasons are that the experience is short, that disciplines are young (even if spreadsheet programs have been existing for years). There is also a big temptation to focus on products and concrete results rather than on real abilities and competences to develop among learners. In this paper, we suggested a method to elicit fundamental knowledge related to the spreadsheet program. This method is based on the common existing features of all the spreadsheet programs and the concepts they convey. We named these common features "principles". And we named principles and concepts "invariants".

Invariants help us to define a trail to discover the power of the spreadsheet programs, but they are also means, a way of thinking that helps learners to go further by themselves. Notice that invariants may be identified for each software package, not only for the spreadsheet program.

Once knowledge elicited, the design of the trail will depend on the audience. Here, we incompletely illustrated an example with some flashes on a particular trail, trying to precise some of its specific objectives. A few examples were given to suggest some recommendations like providing illustrations and exercises, that…
- focus on the main concept(s) (*i.e.* put the challenge on what has to be learned);
- promote exploration (*i.e.* use of a particular function, discovery of how the system works);
- excite learners' motivation through challenges.

All this looks very simple. But choosing good examples, creating interesting exercises, respecting the three previous recommendations take time. If you find such examples, keep them carefully and better, try to share them with others. Books about spreadsheets are full of illustrations showing sales tables, bills and pay slips or whatever, but examples like those we showed remain unfortunately rare.